# Deformation and tearing of graphene-reinforced elastomer nanocomposites

Mufeng Liu[1], Jason H. Hui[1], Ian A. Kinloch[1], Robert J. Young[1*], Dimitrios G. Papageorgiou[2*]

[1]National Graphene Institute, Henry Royce Institute and Department of Materials, University of Manchester, Oxford Road, Manchester M13 9PL, UK
[2]School of Engineering and Materials Science, Queen Mary University of London, Mile End Road, London E1 4NS, UK
*Corresponding authors: d.papageorgiou@qmul.ac.uk, robert.young@manchester.ac.uk*

## Abstract

The resistance to failure through tearing is a crucial mechanical property for the application of different elastomers. In this work, graphene nanoplatelets (GNPs) were introduced into a fluoroelastomer (FKM) matrix with the aim of improving its tear resistance. The fracture energy through tearing was evaluated using the pure shear test. It was found that the tearing energy increased linearly with the volume fraction of the GNPs. At the maximum GNP content, the tearing resistance was 3 times higher, suggesting efficient toughening from the GNPs. Theoretical analysis of the micromechanics was conducted by considering debonding and pull-out of the nanoplatelets as possible toughening mechanisms. It was determined quantitatively that the main toughening mechanism was debonding of the interface rather than pull-out. The formation of cavities at flake ends during the deformation, as confirmed by scanning electron microscopy, was found to contribute to the remarkably high interfacial debonding energy (~1 kJ/m$^2$).

## 1. Introduction

Fluoroelastomer (FKMs) systems are used widely in the aerospace and automotive industries due to primarily their excellent barrier properties against aggressive non-polar chemicals [1]. Most of the FKM materials that are used in sealing components operate under elevated temperatures, usually also bearing static forces [1]. It is therefore inevitable that defects can be generated within the materials (such as the result of liquid or gas molecules diffused in the materials that end up initiating cavities), while external forces can induce internal cracks, which are followed by tear opening, propagation and eventual failure of materials [2]. It is therefore important to improve the tearing resistance of FKM components by, for example, the introduction of high-performance inorganic fillers such as graphene-related materials (GRMs).

It is well-accepted that graphene-based fillers can improve multifunctional properties of rubbers due to their excellent intrinsic properties and high aspect ratios. [3-7]. In addition, it has been shown that the tear resistance of elastomers can also be enhanced by introducing GRMs [8-11]. This is particularly significant, since tear resistance is a critical parameter for numerous elastomer applications, although the mechanisms of toughening remain unclear. Over the years, the improvement of tear strength by inorganic micro- and nano-fillers, including glass beads [12], clays [13, 14] and GRMs [9-11], has been explained mainly in terms of a deviation of the tear path (where the fillers act as a physical barrier to crack propagation) [12, 15]. Such an explanation was believed to be valid for micron-sized fillers, as it was found that large sized particles (at least hundreds of microns in every dimension) can increase the roughness of the tear surface [12]. However, even in the 1980s, before the emergence of polymer nanocomposites, Dreyfuss *et al.* pointed out that nano-fillers of small particle size have little influence on the roughness of the tear surfaces of elastomers [12].

In the present study, we have investigated the mechanisms of the tearing in a graphene-reinforced FKM. The FKM/graphene nanoplatelet (GNP) nanocomposites were prepared using solid-state mixing in a two-roll mill and were subsequently moulded by hot pressing. Three main methods are usually used to evaluate the tear resistance of rubbers, i.e., I. the ASTM D624, type C, using an un-nicked test piece with a 90° angle on one side and with tab ends; II. simple extension tearing (trouser-shaped samples) [16] and III. pure shear testing [16, 17]. We performed pure shear tests on the FKM/GNP nanocomposites, which enabled us to obtain a valid tearing energy for the samples [17]. The microstructure of the tear surfaces was examined using scanning electron microscopy (SEM). The tearing energy of the nanocomposites was modelled as a function of the volume fraction of the graphene nanoplatelets using fracture mechanics, and this enabled us to elucidate the mechanisms leading to the improvement of tear resistance as a result of the addition of the GNPs.

## 2. Experimental Procedure

### 2.1 Materials and processing

The processing procedure is summarised in the flow chart shown in Figure S1. The fluoroelastomer (FKM), Tecnoflon® PFR 94 (Solvay Ltd.) was used as received. The graphene nanoplatelets (GNPs), received from Avanzare Ltd., coded as AVA-0240, with a specific surface area (BET) of $37 \pm 12$ m$^2$/g [33], were mixed with the FKM. The other additives employed in rubber processing for more effective mixing and curing, peroxide (Luperox® 101XL45, Arkema Co. Ltd.) and triallyl isocyanurate (TAIC 50 Co-activator, Wilfrid Smith, Ltd.) were of analytical grade and used as received. The detailed formulations of the samples produced in this work are listed in Table S1.

The FKM was compounded with the AVA-240 at nominal loadings of 2.5, 5, 10 and 15 phr (parts by weight per hundred parts of rubber) in a two-roll mill at ambient temperature. The prepared rubber compounds were then cut into pieces and hot pressed into sheets (~2.5 mm thick) using a Collin Platen Press (Platen Press P 300 P/M). The vulcanization took place at a temperature of 160 °C for 10 minutes under a hydraulic pressure of 30 bar. Afterwards, the hot-pressed sheets were post-cured in an oven at 230 °C for 2 hours. This method of processing led to the majority of the GNPs being aligned in the plane of the sheets. All the moulded elastomer sheets (~2.5 mm thick) were then cut into tear testing samples with pre-cut cracks for the pure shear tests [17].

### 2.2 Characterisation

The actual loadings of the fillers in the nanocomposites were determined by thermogravimetric analysis (TGA) using a Jupiter® thermal analyzer (Netzsch STA 449 F5). The samples were heated from room temperature up to 600 °C under a 50 mL/min flow of N$_2$ at 10°C/min. At least 3 measurements were performed for each sample. The results of the actual GNP loadings and the translation to filler volume fractions are described in S1 and tabulated in Table S2.

Tear testing was carried out using an Instron 3365 with a load cell of 5 kN. At least 5 specimen were tested for each sample. The tensile tests were undertaken using a cross-head speed of 50 mm·min$^{-1}$ in accordance with the pure shear measurement protocol proposed by Rivlin and Thomas [17]. Tensile testing was also conducted based on ASTM D412 to evaluate the tensile modulus of the samples.

The fracture surfaces of the nanocomposites after the tear tests were examined using scanning electron microscopy (SEM). The samples were coated with 3 nm of Au/Pd alloy and the SEM micrographs were obtained using a TESCAN Mira 3 Field Emission Gun Scanning Electron Microscope (FEGSEM) operated at 5 kV.

## 3. Result and Discussion

### 3.1 Tear resistance

The tear resistance of the FKM and nanocomposite samples was evaluated by using the pure shear tear test protocol as described in the seminal work of Rivlin and Thomas [17] which enables valid fracture mechanics data to be determined [18]. As can be seen from the schematic diagram of a stress-extension curve in **Figure 1**(a), with the application of the stress, the materials underwent tensile deformation and the pre-cut crack was deformed and took the form of a semi-circle at the tip of the crack. When the stress reached its maximum point, the tearing initiation stress (TIS) was reached, followed by tear propagation until the material fractured completely. The tearing energy can be then calculated by integration of the area under the stress strain curve from the TIS point up to the failure point [17], as shown in **Figure 1**(a).

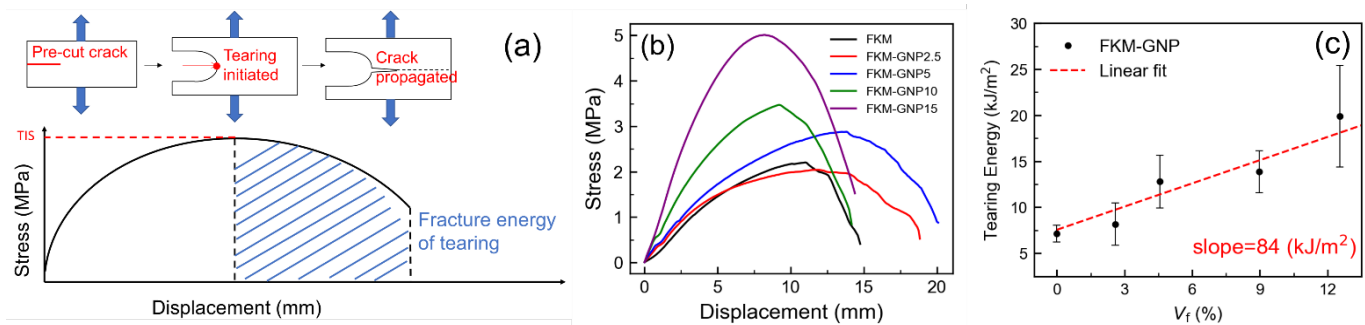

**Figure 1**. (a) Schematic diagram of a stress-displacement curve for a specimen under pure shear testing. (b) Representative stress-displacement curves of all specimens from the pure shear test. (c) Calculated tearing energy against the volume fraction of the GNPs.

Typical stress-displacement curves of the different nanocomposite samples subjected to the pure shear test are presented in **Figure 1**(b). It can be seen that the initial slope of the curves increased with increasing GNP loading, indicating the stiffening effect of the GNPs [7]. Additionally, the tear initiation stress (TIS) increased with increasing filler loading. The tearing energy for all samples under study was calculated using the method shown schematically in **Figure 1**(a). The results are plotted in **Figure 1**(c) as a function of the volume fraction of the filler. It can be seen that the tearing energy increased linearly with increasing volume fraction of the nanoplatelets. Compared with neat FKM, at the maximum GNP loading of 15 phr (~12 vol%), the fracture energy of tearing was approximately a factor of three higher, indicating effective reinforcement. The mechanisms of the improvement of the tearing resistance are analysed using micromechanics in the following sections.

## 3.2 Fracture surfaces

The microstructures of the tear surfaces of all the samples are shown in **Figure 2**. The neat FKM in **Figure 2**(a) and Figure S5 (a-b) had a fracture surface without visible cavities/holes. For the nanocomposite samples, it can be seen clearly that nanoplatelet-shaped holes are present in the tear surfaces, resulting from pulled-out flakes. In addition, a large number of small cavities (1 μm or smaller) with more or less spherical shapes are present in the fracture surfaces (see **Figure 2** b-e and Figure S5 c-f), which is evidence of cavitation. These small spherical cavities, however, did not expand to large cracks after tearing, possibly due to lack of adequate stress that would allow them to tear open during the deformation, since it was demonstrated that smaller cavities require higher stress to expand and form visible cracks, particularly when the size of cavities is below one micron [2]. An interesting feature can also be seen in **Figure 2** (b-e) and Figure S5 (c-f), where the interfaces of small flakes (less than 5 μm) remained intact after the tearing as the flakes still remain well-wetted by the polymer, while larger flakes (larger than ~30 μm) show clear debonding as well as the resulting holes. This could be the result of the higher stress and/or energy that is required to induce cavitation for small impurities (of the order of 1 μm) in rubber systems [2].

The average lateral length of the AVA-240 GNP flakes is reported to be 50 μm [19] whereas it can be seen from the SEM images in **Figure 2** that the lateral size of the flakes is significantly smaller than the nominal value reported [19]. Additionally the presence of loops/folds in the reinforcement can also be observed. A lateral length of 50 μm combined with a specific surface area of $37 \pm 12$ m$^2$/g leads to a GNP aspect ratio (length/thickness) of the order of $10^3$. Our previous studies have shown that such high aspect ratios are not realised for mechanical reinforcement and the effective aspect ratio of the GNP flakes in nanocomposites is generally of the order of 100 [3, 5-7, 20-23].

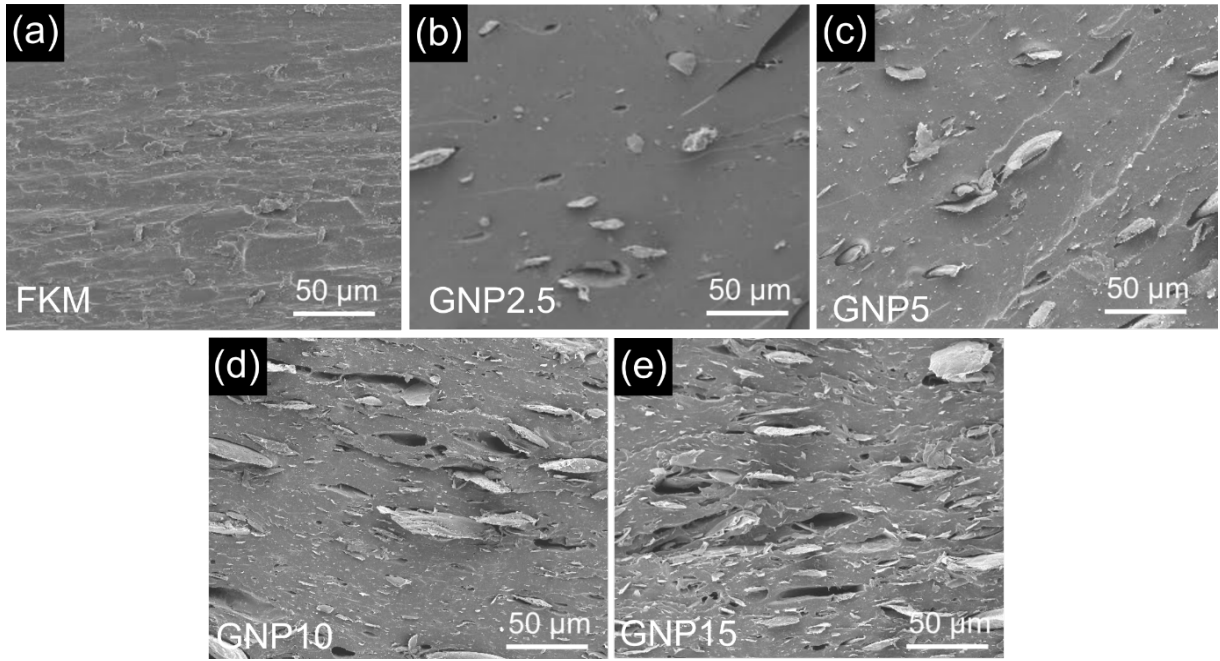

**Figure 2**. SEM micrographs of the fractured surfaces of the samples after tear tests: (a) FKM, (b-e) FKM filled with 2.5, 5, 10 and 15 phr of GNPs. The alignment of the GNPs in the plane of the sheets can be seen.

### 3.3 The mechanics of fracture of the nanocomposites

We have modified a well-established model, developed previously by Hull and Clyne for the determination of the fracture energy of short-fibre composites [24], to analyse the tearing resistance of graphene reinforced rubber nanocomposites from the pure shear tests. The complete derivation of equations for the mechanisms of both debonding and pull-out can be found in the Supporting Information (Section S3).

The analysis of both mechanisms shows that the effective aspect ratio ($s_{eff}$) of the GNPs is an important parameter in both analyses. The value of $s_{eff}$ was determined independently for the FKM/GNP nanocomposite system studied, based on the analysis of the tensile modulus of the samples using the well-defined shear-lag/rule-of-mixtures theory [4, 5, 20] and found to be ~80 (see Section S2 of the Supporting Information). This value of the aspect ratio $s_{eff}$, is lower than the nominal value of around $10^3$ discussed earlier and can be attributed to the looped/folded flake morphologies within the rubber matrix, as well as agglomeration of the GNPs in the nanocomposites.

We can use the slope of the fracture energy of tearing against the volume fraction of the GNPs (**Figure 1**c) to evaluate the fracture energy of debonding and pull-out using the equations that are described in the subsequent section and derived in detail in Section S3 of the Supporting Information. The possible contribution of each mechanism has been considered separately.

*3.3.1 Debonding as the main mechanism of failure*

Assuming that fracture of the flakes did not take place during the deformation of the materials at such low stresses, the failure of the nanocomposites can be attributed to polymer/filler debonding followed by pull-out of the aligned flakes. By considering the work needed for the interfacial debonding of a single flake and expanding this to take into account the total number of flakes and the total work of debonding (S3.2-Supporting Information), the fracture energy for debonding is given by the following equation,

$$G_{cd}=sG_{ic}V_f \qquad (s=l/t) \qquad (1)$$

where $s$ is the aspect ratio of the GNPs, $l$ and $t$ are the length and the thickness of the flakes; $G_{ic}$ is the fracture energy of the interface (per unit area of the interface). If it is assumed that debonding is the main mechanism controlling the fracture of the nanocomposite, the interfacial fracture energy per unit area, $G_{ic}$, can be determined to be ~1 kJ/m² using equation (1), for an effective aspect ratio of the flakes of 80 (the calculated values of $G_{ic}$ for each GNP loading were tabulated in Table S3). Equation (1) shows that a higher aspect ratio would lead to a lower value of $G_{ic}$.

Gent *et al.* [25] studied the pull-out of metallic rods from a transparent rubber block and determined that the fracture energy for interfacial debonding in the rubber-based system was in the order of ~300 J/m², due to the formation and expansion of cavities at the embedded fibre end. This value of 0.3 kJ/m² was obtained for a large rod (in the order of 1 mm in diameter) pulled out from a rubber block and it was demonstrated that the debonding energy increased with a decreasing radius of the rods [25]. For the FKM/GNP nanocomposites, the lateral size of the GNPs is in the order of 10s of microns [19] (Figure 2) consistent with a higher debonding energy (~1 kJ/m²).

In order to adapt the cavitation concept for elastomers reinforced by 2D nanofillers, a schematic diagram of the process of formation and expansion of the cavities is presented schematically in **Figure 3**. During the pull-out of the flake, a small cavity may form initially at the flake ends (preferably at a corner of a flake based on Gent's findings [25] - left of **Figure 3**). The cavity then expands (middle of **Figure 3**) and eventually a complete separation of the flake end takes place (right of **Figure 3**). It should be pointed out that the $G_{ic}$ value reported herein is significantly higher than the debonding energy for typical fibre/epoxy composites, which is only of the order of 10 J/m² [24]. This is the result of the different failure mechanism of elastomers - cavitation - compared to epoxy resins where cavitation is not found to take place [2, 25, 26]. The interfacial interactions between the GNPs and the matrix play a major role in tear resistance [27-29]. The results presented in this work have indicated excellent interfacial interactions between the GNP and the FKM matrix, given that the tearing energy increased linearly with increasing filler content. In the case of poor filler/matrix interactions, the GNPs and the surrounding interface should be approximated to voids which reduce the tearing energy and the tensile modulus of the materials; this is clearly not observed herein.

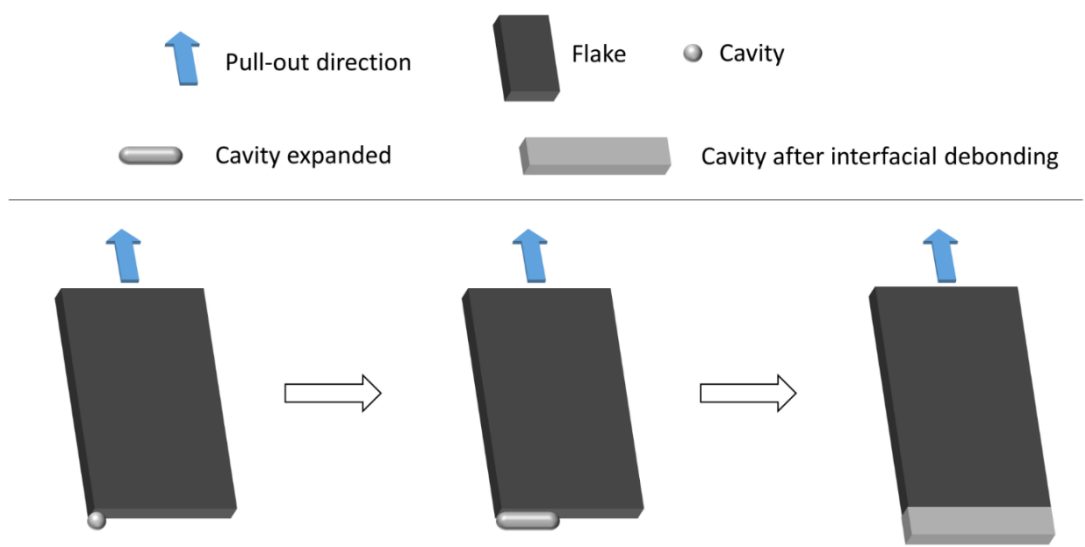

**Figure 3**. Schematic diagram of the formation (left) of cavity during pull-out, cavity expansion (middle) and cavity after interfacial debonding (right).

### 3.3.2 Flake pull-out as the main mechanism of failure

If flake pull out rather than debonding is considered to be the main contribution to the fracture energy, the energy required to pull the flakes out of the holes after debonding (see Section S2.3 in the Supporting Information) is given by,

$$G_{cp} = s\frac{l\tau_i^*}{12}V_f \qquad (s=l/t) \qquad (2)$$

where $l$ is the length of the flake and $\tau_i^*$ is the interfacial shear stress, respectively.

Assuming that pull-out/sliding is the main energy adsorption mechanism during fracture, the interfacial shear stress $\tau_i^*$, is calculated to be 250 MPa using equation (2), for an effective aspect ratio of the flakes of 80, and a flake length of 50 μm for AVA-240 graphene nanoplatelets [19] (the calculated values of $\tau_i^*$ for each GNP loading were tabulated in Table S4). It is shown in Section S3, Figure S2(b) that the tensile modulus of the FKM at 50% strain is only around 1.5 MPa (shear modulus ~0.5 MPa). Hence it is very unlikely that such a large interfacial shear stress could be sustained during pull-out. Equation (2) shows that the value of interfacial shear stress would be even higher for a flake length lower than 50 μm.

The predictions for both the mechanics of debonding and pull-out are that the fracture energy is proportional to the volume fraction of the GNPs, consistent with the experimental results obtained herein. Also, according to different literature reports, in graphene/elastomer nanocomposites, the tear strength of the nanocomposites has a linear relationship with the filler volume fraction [27, 30, 31]. We have demonstrated that the different mechanisms can only be distinguished using the analysis outlined in the Supporting Information, where the mechanics of each process are considered in detail.

From the above, although it is clear that the processes of debonding and pull-out both take place during tearing of the GNP/FKM nanocomposite, it can be concluded that energy needed for debonding contributes dominantly to the increased fracture energy of the tearing of the FKM/GNP nanocomposites. This is most likely due to the debonding of the FKM/GNP interfaces taking place via a process that involves cavitation at the nanoplatelet ends. In contrast the process of pull-out would appear unlikely to absorb much energy. This behaviour needs to be contrasted with the fracture of short-fibre reinforced composites with rigid matrices [24] where the bulk of the energy adsorption is thought to be through frictional sliding rather than debonding.

## 4. Conclusions

The tearing resistance of the fabricated FKM/GNP nanocomposites was evaluated using the pure shear test. It was found that the fracture energy of tearing increased linearly with increasing volume fraction of the GNPs. At the maximum loading of 15 phr of GNPs in the FKM, the tearing energy was found to be three times higher than that of the matrix, indicating effective toughening through the addition of the GNPs. Theoretical analyses were conducted using micromechanics, based upon the mechanisms of debonding and pull-out/sliding, that have been used in the past to analyse typical fibre/epoxy composites. Both mechanisms were predicted to lead to a linear relationship between the fracture energy and the volume fraction of the GNPs. It was determined quantitatively that debonding of the FKM/GNP was most likely the main energy adsorption mechanism. The fracture energy of interfacial debonding was estimated to be 1 kJ/m$^2$, significantly higher than that for conventional fibre/epoxy composites, due to cavitation phenomena at the edge of graphene nanoplatelets in the graphene-elastomer nanocomposites. It was revealed from the microstructural fractography analysis that smaller size flakes (a few of microns) tended to produce intact interfaces after tearing, while larger flakes (in tens of microns) show failed interfaces and visible holes as a result of flake pull-out. This could be due to the energy being required for the expansion of smaller cavities at the interfaces of smaller flakes.

## Acknowledgements

The authors acknowledge the support from "Graphene Core 2", GA: 785219 and "Graphene Core 3" GA: 881603 which are implemented under the EU-Horizon 2020 Research & Innovation Actions (RIA) and are financially supported by EC-financed parts of the Graphene Flagship. Ian A. Kinloch also acknowledges the Royal Academy of Engineering and Morgan Advanced Materials. All research data supporting this publication are available within this publication.

# References

[1] A.L. Moore, 1 - Fundamentals, in: A.L. Moore (Ed.), Fluoroelastomers Handbook, William Andrew Publishing, Norwich, NY, 2006, pp. 3-12.

[2] A.N. Gent, C. Wang, Fracture mechanics and cavitation in rubber-like solids, Journal of Materials Science 26(12) (1991) 3392-3395.

[3] S. Li, Z. Li, T.L. Burnett, T.J.A. Slater, T. Hashimoto, R.J. Young, Nanocomposites of graphene nanoplatelets in natural rubber: microstructure and mechanisms of reinforcement, Journal of Materials Science 52(16) (2017) 9558-9572.

[4] R.J. Young, M. Liu, I.A. Kinloch, S. Li, X. Zhao, C. Vallés, D.G. Papageorgiou, The mechanics of reinforcement of polymers by graphene nanoplatelets, Composites Science and Technology 154 (2018) 110-116.

[5] M. Liu, D.G. Papageorgiou, S. Li, K. Lin, I.A. Kinloch, R.J. Young, Micromechanics of reinforcement of a graphene-based thermoplastic elastomer nanocomposite, Composites Part A: Applied Science and Manufacturing 110 (2018) 84-92.

[6] D.G. Papageorgiou, I.A. Kinloch, R.J. Young, Mechanical properties of graphene and graphene-based nanocomposites, Progress in Materials Science 90 (2017) 75-127.

[7] D.G. Papageorgiou, I.A. Kinloch, R.J. Young, Graphene/elastomer nanocomposites, Carbon 95 (2015) 460-484.

[8] B. Dong, C. Liu, L. Zhang, Y. Wu, Preparation, fracture, and fatigue of exfoliated graphene oxide/natural rubber composites, Rsc Advances 5(22) (2015) 17140-17148.

[9] Y. Mao, S. Wen, Y. Chen, F. Zhang, P. Panine, T.W. Chan, L. Zhang, Y. Liang, L. Liu, High Performance Graphene Oxide Based Rubber Composites, Scientific Reports 3(1) (2013) 2508.

[10] X. Liu, W. Kuang, B. Guo, Preparation of rubber/graphene oxide composites with in-situ interfacial design, Polymer 56 (2015) 553-562.

[11] L. Zheng, S. Jerrams, Z. Xu, L. Zhang, L. Liu, S. Wen, Enhanced gas barrier properties of graphene oxide/rubber composites with strong interfaces constructed by graphene oxide and sulfur, Chemical Engineering Journal 383 (2020) 123100.

[12] P. Dreyfuss, A. Gent, J. Williams, Tear strength and tensile strength of model filled elastomers, Journal of Polymer Science: Polymer Physics Edition 18(10) (1980) 2135-2142.

[13] L. Zhang, Y. Wang, Y. Wang, Y. Sui, D. Yu, Morphology and mechanical properties of clay/styrene‐butadiene rubber nanocomposites, Journal of Applied Polymer Science 78(11) (2000) 1873-1878.

[14] H. Zheng, Y. Zhang, Z. Peng, Y. Zhang, Influence of clay modification on the structure and mechanical properties of EPDM/montmorillonite nanocomposites, Polymer Testing 23(2) (2004) 217-223.

[15] A. Mohammad, G.P. Simon, 12 - Rubber-clay nanocomposites, in: Y.-W. Mai, Z.-Z. Yu (Eds.), Polymer Nanocomposites, Woodhead Publishing2006, pp. 297-325.

[16] A. Thomas, Rupture of rubber. VI. Further experiments on the tear criterion, Journal of Applied Polymer Science 3(8) (1960) 168-174.

[17] R.S. Rivlin, A.G. Thomas, Rupture of rubber. I. Characteristic energy for tearing, Journal of Polymer Science 10(3) (1953) 291-318.

[18] A.J. Kinloch, R.J. Young, Fracture Behaviour of Polymers, Elsevier1985.

[19] A.C. Ferrari, U.H. Friedrich, T.B. Pokroy, Graphene Core 1 Graphene-Based Disruptive Technologies Horizon 2020 RIA.

[20] M. Liu, P. Cataldi, R.J. Young, D.G. Papageorgiou, I.A. Kinloch, High-performance fluoroelastomer-graphene nanocomposites for advanced sealing applications, Composites Science and Technology 202 (2021) 108592.

[21] M. Liu, I.A. Kinloch, R.J. Young, D.G. Papageorgiou, Modelling mechanical percolation in graphene-reinforced elastomer nanocomposites, Composites Part B: Engineering 178 (2019) 107506.

[22] M. Liu, I.A. Kinloch, R.J. Young, D.G. Papageorgiou, Realising biaxial reinforcement via orientation-induced anisotropic swelling in graphene-based elastomers, Nanoscale 12(5) (2020) 3377-3386.

[23] M. Liu, S. Li, I.A. Kinloch, R.J. Young, D.G. Papageorgiou, Anisotropic swelling of elastomers filled with aligned 2D materials, 2D Materials 7(2) (2020) 025031.
[24] D. Hull, T. Clyne, An Introduction to Composite Materials, Cambridge University Press, Cambridge, 1996.
[25] A. Gent, C. Shambarger, Pull-out of short rods and fibres, Journal of materials science 29(8) (1994) 2107-2114.
[26] K. Cho, A. Gent, Cavitation in model elastomeric composites, Journal of materials science 23(1) (1988) 141-144.
[27] H. Kang, K. Zuo, Z. Wang, L. Zhang, L. Liu, B. Guo, Using a green method to develop graphene oxide/elastomers nanocomposites with combination of high barrier and mechanical performance, Composites Science and Technology 92 (2014) 1-8.
[28] H. Qin, C. Deng, S. Lu, Y. Yang, G. Guan, Z. Liu, Q. Yu, Enhanced mechanical property, thermal and electrical conductivity of natural rubber/graphene nanosheets nanocomposites, Polymer Composites 41(4) (2020) 1299-1309.
[29] M. Du, B. Guo, Y. Lei, M. Liu, D. Jia, Carboxylated butadiene–styrene rubber/halloysite nanotube nanocomposites: Interfacial interaction and performance, Polymer 49(22) (2008) 4871-4876.
[30] S. Araby, Q. Meng, L. Zhang, H. Kang, P. Majewski, Y. Tang, J. Ma, Electrically and thermally conductive elastomer/graphene nanocomposites by solution mixing, Polymer 55(1) (2014) 201-210.
[31] Z. Yang, J. Liu, R. Liao, G. Yang, X. Wu, Z. Tang, B. Guo, L. Zhang, Y. Ma, Q. Nie, F. Wang, Rational design of covalent interfaces for graphene/elastomer nanocomposites, Composites Science and Technology 132 (2016) 68-75.

# Deformation and tearing of graphene-reinforced elastomer nanocomposites

*Mufeng Liu[1], Jason H. Hui[1], Ian A. Kinloch[1], Robert J. Young[1*], Dimitrios G. Papageorgiou[2*]*

[1]*National Graphene Institute, Henry Royce Institute and Department of Materials, University of Manchester, Oxford Road, Manchester M13 9PL, UK*

[2]*School of Engineering and Materials Science, Queen Mary University of London, Mile End Road, London E1 4NS, UK*

*Corresponding authors: d.papageorgiou@qmul.ac.uk, robert.young@manchester.ac.uk*

## *Supporting Information*

### *S1. Primary Information of the FKM nanocomposites*

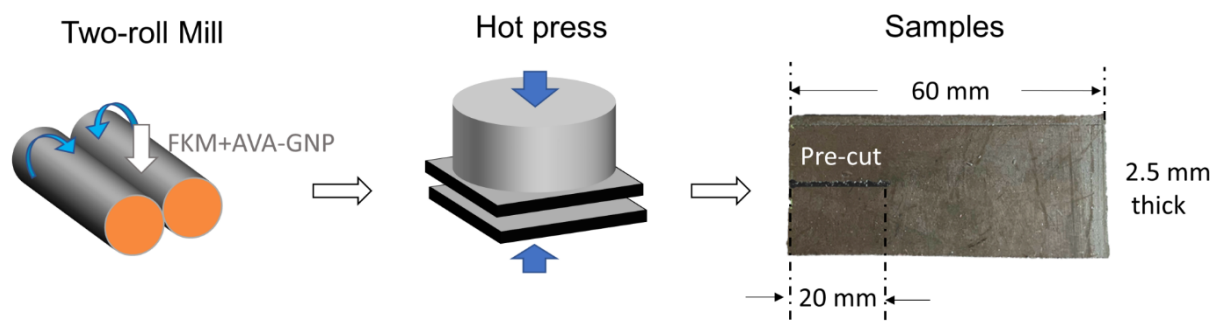

*Figure S1. Flow chart of the processing method from two-roll mill to hot press to sample dimensions.*

Table S1. Formulation of the FKM compounds

| Materials | Loading (phr*) |
|---|---|
| FKM (Tecnoflon® PFR 94) | 100 |
| Peroxide (Luperox® 101XL45) | 1.5 |
| TAIC 50 Co-activator | 6 |
| GNP (AVA-0240) | 0, 2.5, 5, 10, 15 |

*phr stands for part per hundred rubber by mass

The transformations from the mass fraction to the volume fraction of the fillers in the nanocomposites were achieved using equation 1, with the densities of GNP (2.2 g/cm$^3$), CB (2.0 g/cm$^3$) and FKM (2.4 g/cm$^3$) provided by the suppliers.

$$V_f = \frac{w_f \rho_m}{w_f \rho_m + (1-w_f)\rho_f} \qquad (S1)$$

where $w_f$ is the weight percentage obtained from TGA, $\rho_m$ and $\rho_f$ are the density of the matrix and the fillers, respectively.

Table S2. Mass residues and mass fractions of the fillers obtained by thermogravimetric analysis (TGA) and volume fractions of the fillers calculated using equation S1.

| phr | Mass residue (%) | Mass fraction (%) | Volume fraction (%) |
|---|---|---|---|
| 0 | 8.30 ± 0.05 | 0 | 0 |
| 2.5 | 10.68 ± 0.10 | 2.38 ± 0.15 | 2.59 ± 0.17 |
| 5 | 12.51 ± 0.09 | 4.21 ± 0.14 | 4.57 ± 0.14 |
| 10 | 16.58 ± 0.15 | 8.28 ± 0.20 | 8.97 ± 0.22 |
| 15 | 19.93 ± 0.05 | 11.63 ± 0.10 | 12.55 ± 0.13 |

## S2. Results of tensile properties and modelling of modulus

The results from tensile testing can be seen in Figure S2. It was shown that the 50 % modulus of the samples can be fitted using shear-lag/rule-of-mixtures theory [1, 2], giving an effective aspect ratio of 80. The theory leads to the following relationship,

$$E_c = E_m(1 - V_f + 0.056 s_{eff}^2 V_f^2) \qquad (S2)$$

where $E_c$ and $E_m$ are the modulus of the nanocomposites and the matrix, respectively; $V_f$ is the volume fraction of the filler and $s_{eff}$ is the effective aspect ratio of the graphene nanoplatelets [1-3].

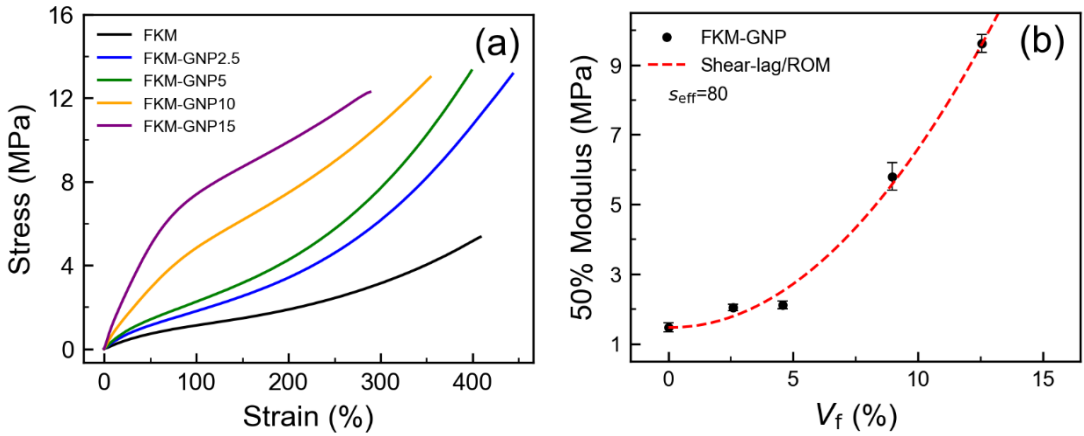

Figure S2. (a) Typical stress-strain curves of FKM and its nanocomposites under tensile deformation; (b) 50 % modulus of the FKM and its nanocomposites with curve fitting of shear-lag/rule-of-mixtures theory. It was shown that the effective aspect ratio of the flakes obtained from the fitting is 80.

## S3. Modelling of Crack Propagation in Graphene-reinforced FKM Nanocomposites

The mechanics of crack propagation of FKM/GNP nanocomposites was adapted from the analysis of the fracture behaviour of short-fibre composites [4]. Assuming that fracture of the flakes did not take place, the failure of the FKM/GNP nanocomposites is assumed to involve a combination of debonding followed by pull-out of the flakes.

### S3.1. Model flake and debonding geometry

The model flake is shown in Figure S2. The lateral size and the thickness of the model flake are $l$ and $t$ respectively. The aspect ratio of the flake is given by the ratio of the lateral size over the thickness of the flake, $s = l/t$.

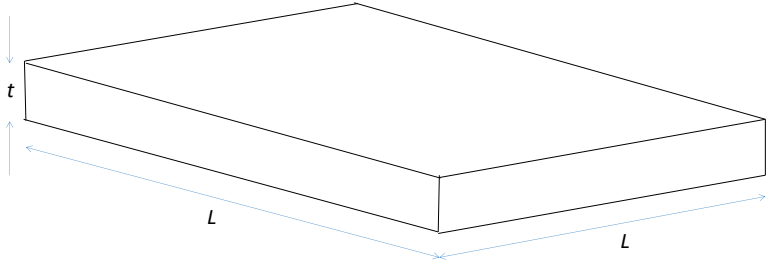

*Figure S3. Model flake with a lateral size of l and a thickness of t.*

*The debonding geometry is shown in Figure S4. A length of x was set to be the embedded length of the flake before debonding. After debonding, the length increased to $x_0$ and was defined as the debonding distance, $x_0 \leq l/2$.*

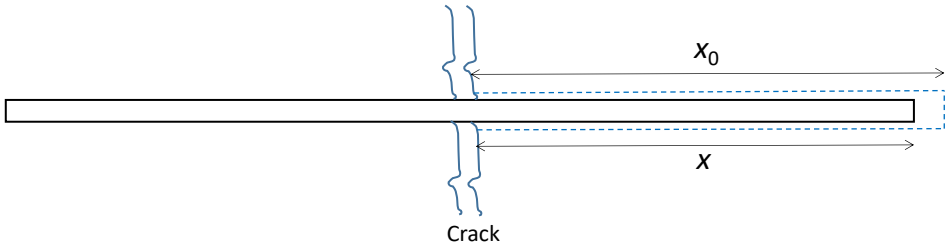

*Figure S4. Debonding geometry of a graphene flake in FKM, with a debonding distance of $x_0$ and the embedding length of x before debonding, where $x_0 \leq l/2$.*

### S3.2 Mechanics of Debonding

*The work done ($\Delta U$) for the interfacial debonding of a single flake is given by,*

$$\Delta U = (2l + 2t) x_0 G_{ic} \quad (S3)$$

*where $G_{ic}$ is the fracture energy of the interface (per unit area of the interface).*

*The total number of flakes with an embedded length of between $x_0$ and $(x_0 + dx_0)$ is,*

$$N dx_0 / (l/2) = 2 N dx_0 / l \quad (S4)$$

*where N is the number of flakes per unit area and N can be correlated to the volume fraction of the flakes ($V_f$) and flake geometry,*

$$N = V_f / (\text{cross-sectional area of one flake}) = V_f / lt \quad (S5)$$

*The total work of debonding is given by,*

$$G_{cd}= \int_0^{\frac{l}{2}} \text{(number of flakes)} \times \text{(work for debonding a single flake)} \qquad (S6)$$

$$G_{cd}= \int_0^{\frac{l}{2}} \frac{2V_f}{l^2 t} dx_0 \, 2(l+t)x_0 G_{ic} \qquad (S7)$$

*Since $l \gg t$, for a graphene flake,*

$$G_{cd} \approx \int_0^{\frac{l}{2}} \frac{4V_f}{lt} G_{ic} x_0 dx_0 \qquad (S8)$$

$$G_{cd}= \frac{4V_f}{lt} \frac{l^2}{4} G_{ic} = sG_{ic}V_f \qquad (s = l/t) \qquad (S9)$$

*It can be concluded that the fracture energy for debonding $G_{cd}$, is predicted to be proportional to both the volume fraction of flakes and their aspect ratio.*

Table S3. Calculated fracture energy of the interface ($G_{ic}$) for each filler contents based on the assumption of 3.3.1/S3.2 and equation (1)/ (S9), corresponding to the data for the linear curve fitting in Figure 1(c) in main text.

| phr | Volume fraction (%) | Tearing energy, $G_{cd}$ (kJ/m$^2$) | Fracture energy of the interface, $G_{ic}$ (kJ/m$^2$) |
|---|---|---|---|
| 0 | 0 | 7.15 ± 0.91 | - |
| 2.5 | 2.59 ± 0.17 | 8.18 ± 2.32 | 0.50 ± 0.68 |
| 5 | 4.57 ± 0.14 | 12.79 ± 2.84 | 1.54 ± 0.53 |
| 10 | 8.97 ± 0.22 | 13.87 ± 2.28 | 0.94 ± 0.19 |
| 15 | 12.55 ± 0.13 | 19.91 ± 5.51 | 1.27 ± 0.46 |

### S3.3 Mechanics of pull-out/frictional sliding

The energy required to pull the flakes out of the holes after debonding should be taken into consideration. The work done on a single flake is given by the product of the force acting on the interface multiplying the distance moved,

$$dU = (2l+2t) x \tau_i^* dx \quad (S10)$$

where $\tau_i^*$ is the interfacial shear stress.

The work done in pulling out the flake completely is given by,

$$\Delta U = \int_0^{x_0} 2(l+t) x \tau_i^* dx \quad (S11)$$

and therefore,

$$\Delta U = (l+t) x_0^2 \tau_i^* \quad (S12)$$

Integrating over all flakes being pulled out, the pull-out energy is given by,

$$G_{cp} = \int_0^{l/2} (\text{number of flakes}) \times (\text{work done pulling out a single flake}) \quad (S13)$$

$$G_{cp} = \int_0^{\frac{l}{2}} \frac{2V_f}{l^2 t} dx_0 (l+t) x_0^2 \tau_i^* \qquad (S14)$$

*Since $l \gg t$, for a graphene flake,*

$$G_{cp} \approx \int_0^{\frac{L}{2}} \frac{2V_f}{lt} \tau_i^* x_0^2 dx_0 \qquad (S15)$$

$$G_{cp} = s \frac{l \tau_i^*}{12} V_f \qquad (S16)$$

*In conclusion, the fracture energy for sliding/pull-out is also predicted to be proportional to both the volume fraction of flakes and their aspect ratio.*

*Table S4. Calculated interfacial shear stress ($\tau_i^*$) for each filler content based on the assumption of 3.3.2/S3.3 and equation (2)/(S16), corresponding to the data for the linear curve fitting in Figure 1(c) in main text.*

| phr | Volume fraction (%) | Energy of pull-out, $G_{cp}$ (kJ/m$^2$) | Interfacial shear stress, $\tau_i^*$ (MPa) |
|---|---|---|---|
| 0 | 0 | 7.15 ± 0.91 | - |
| 2.5 | 2.59 ± 0.17 | 8.18 ± 2.32 | 118.91 ± 163.44 |
| 5 | 4.57 ± 0.14 | 12.79 ± 2.84 | 370.24 ± 127.00 |
| 10 | 8.97 ± 0.22 | 13.87 ± 2.28 | 224.75 ± 45.69 |
| 15 | 12.55 ± 0.13 | 19.91 ± 5.51 | 304.90 ± 109.92 |

## *S4. Additional SEM micrographs*

*More SEM micrographs are present in Figure S4, in order to show absence of cavities of the tearing surfaces of the neat FKM samples (a-b) and visible cavities in spherical shape (c-f, highlighted) in FKM/GNP nanocomposite samples. It is also clear in Figure S4(c-f) that small flakes showed intact interface after tearing took place. Such features indicate that the cavitation and expansion of the cavities of the FKM/GNP nanocomposites samples might have happened during the tearing tests, which contributed to higher fracture energy.*

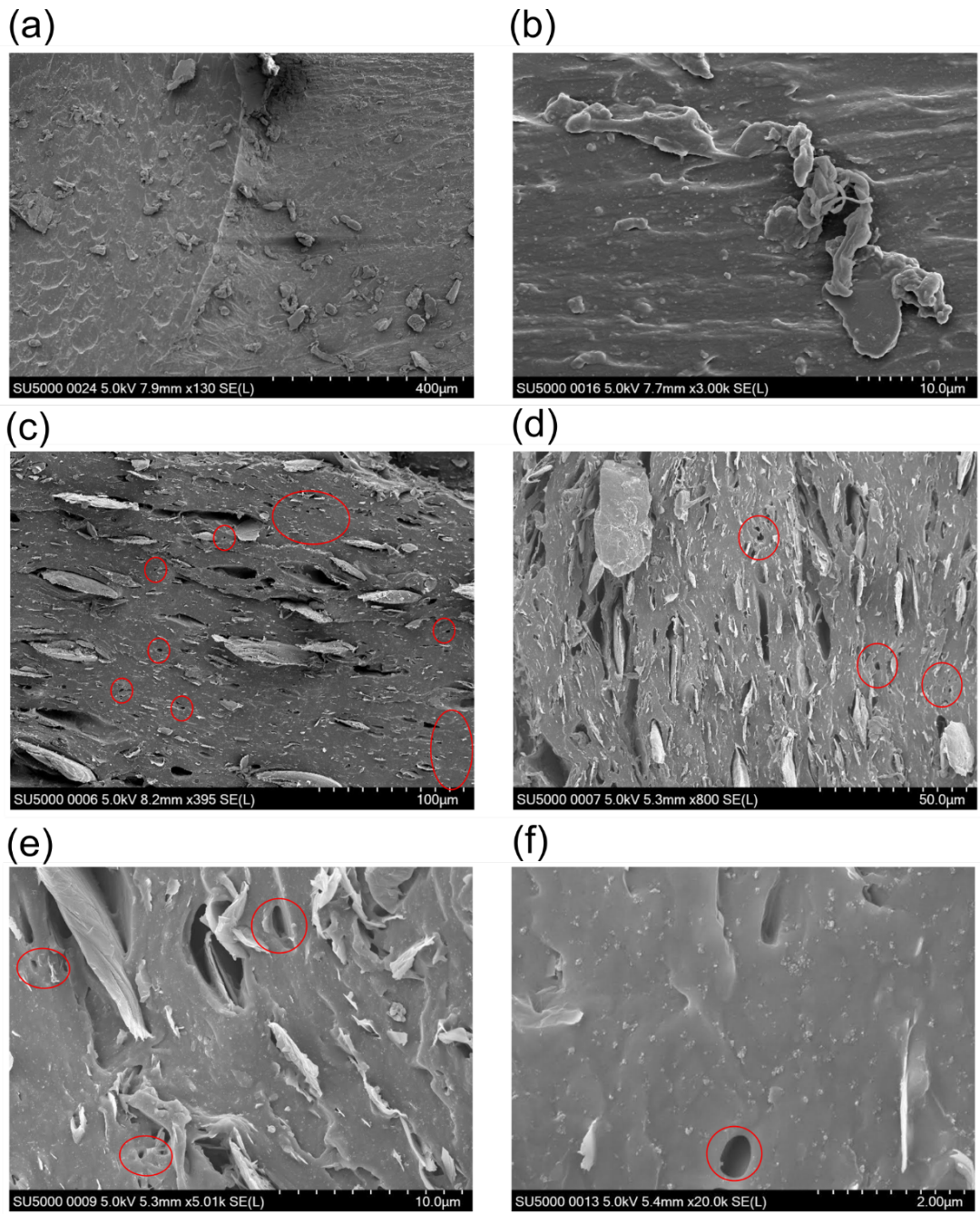

*Figure S5. SEM micrographs of (a,b) low and high magnification images of neat FKM tearing surfaces, where no visible holes or cavities was shown; (c-f) SEM images of tearing surfaces of FKM/GNP nanocomposites, where spherical holes (highlighted) and holes with the configuration of pull-out flakes can both be seen, indicating the cavitation and expansion of cavities might take place during the tearing tests; (c) is FKM-GNP 10 phr and (d-f) are from FKM-GNP 15 phr; It can also be found that the filler/matrix interfaces of small flakes remained intact after tearing.*

## References

[1] M. Liu, D.G. Papageorgiou, S. Li, K. Lin, I.A. Kinloch, R.J. Young, Micromechanics of reinforcement of a graphene-based thermoplastic elastomer nanocomposite, Composites Part A: Applied Science and Manufacturing 110 (2018) 84-92.

*[2] R.J. Young, M. Liu, I.A. Kinloch, S. Li, X. Zhao, C. Vallés, D.G. Papageorgiou, The mechanics of reinforcement of polymers by graphene nanoplatelets, Composites Science and Technology 154 (2018) 110-116.*
*[3] M. Liu, P. Cataldi, R.J. Young, D.G. Papageorgiou, I.A. Kinloch, High-performance fluoroelastomer-graphene nanocomposites for advanced sealing applications, Composites Science and Technology 202 (2021) 108592.*
*[4] D. Hull, T. Clyne, An Introduction to Composite Materials, Cambridge University Press, Cambridge, 1996.*